\pdfoutput=1
\newcommand*{\ATLASLATEXPATH}{}
\documentclass[PAPER, atlasdraft=false, cernpreprint, texlive=2016, UKenglish]{\ATLASLATEXPATH atlasdoc}  
 
\usepackage{\ATLASLATEXPATH atlaspackage}
\usepackage{\ATLASLATEXPATH atlasbiblatex}
 
\usepackage{\ATLASLATEXPATH atlasphysics}
 
\graphicspath{{logos/}{figures/}}

\usepackage{multirow,bigdelim}
 
\AtlasTitle{Measurement of $J/\psi$ production in association with a $W^\pm$ boson with $pp$ data at 8 TeV }
 
\AtlasCoverCommentsDeadline{27 September 2019}
\AtlasJournalRef{JHEP 01 (2020) 095}  
\AtlasDOI{10.1007/JHEP01(2020)095} 
\AtlasAbstract{
 
A measurement of the production of a prompt $J/\psi$ meson in association
with a $W^\pm$ boson
with $W^\pm \rightarrow \mu\nu$ and $J/\psi \rightarrow \mu^+ \mu^-$
is presented for $J/\psi$ transverse momenta in the range $8.5$--$150$~\GeV\
and rapidity $|y_{J/\psi}|<2.1$  using ATLAS data recorded in 2012 at the
LHC\@. The data were taken at a proton--proton centre-of-mass energy
of $\sqrt{s}$ = 8~\TeV\ and correspond to an integrated luminosity
of  $20.3~\mathrm{fb}^{-1}$.
The ratio of the prompt $J/\psi$ plus $W^{\pm}$ cross-section to the
inclusive $W^{\pm}$ cross-section is presented as a differential
measurement as a function of $J/\psi$ transverse momenta and
compared with theoretical predictions using different
double-parton-scattering cross-sections.
 
}
 
\author{The ATLAS Collaboration}
 
\AtlasRefCode{BPHY-2015-05}
 
\PreprintIdNumber{CERN-EP-2018-352}

\AtlasCoverSupportingNote{J/Psi production in association with a W boson}{https://cds.cern.ch/record/2025173}

\AtlasCoverAnalysisTeam{Brad Abbott, Miriam Watson, David Bertsche, Stefanos Leontsinis}
 
\AtlasCoverEdBoardMember{Peter Onyisi (EdBoard~Chair), Soeren Prell, Fabrizio Scuri}

\AtlasCoverEgroupEditors{atlas-BPHY-2015-05-editors@cern.ch}
 
\AtlasCoverEgroupEdBoard{atlas-BPHY-2015-05-editorial-board@cern.ch}
\LEcontact{Steve Lloyd s.l.lloyd@qmul.ac.uk}
 
\hypersetup{pdftitle={ATLAS document},pdfauthor={The ATLAS Collaboration}}
 
\begin{document}
 
\maketitle
 
\tableofcontents

\section{Introduction}

The associated production of prompt $J/\psi$ mesons with $W^\pm$ bosons
provides a powerful probe of the charmonium production mechanism in hadronic collisions,
allowing tests of quantum chromodynamics (QCD) at the
boundary between the perturbative and non-perturbative regimes.
The ATLAS Collaboration has previously presented two analyses of $J/\psi$ mesons
produced in conjunction with vector bosons: the associated production of prompt
$J/\psi + W^\pm$ in $\sqrt{s}=7\,\mathrm{\TeV}$ data~\cite{BPHY-2012-06} and the
production of prompt and non-prompt $J/\psi + Z$ in $\sqrt{s}=8\,\mathrm{\TeV}$ data~\cite{BPHY-2014-01}.
This paper presents a new
measurement of the ratio of the cross-section for associated production of
prompt $J/\psi$ + $W^\pm$ to the
inclusive $W^\pm$ production cross-section with $W^{\pm}\rightarrow \mu\nu$ and
$J/\psi \rightarrow \mu^+ \mu^+$ at a centre-of-mass energy of 8~\TeV, exploiting a
four-fold increase in integrated
luminosity over the previous measurement~\cite{BPHY-2012-06}.  The analysis strategy
closely follows the methods of the earlier papers.
Prompt production refers to a $J/\psi$ meson that is produced directly in the proton-proton collision or indirectly from a heavier charmonium state, while non-prompt production occurs when the $J/\psi$ meson is produced in the decay of a $b$-hadron.
The $J/\psi$ events that are produced from radiative decays of
heavier charmonium states (such as $\chi_c \to \gamma J/\psi$) are not distinguished from directly
produced $J/\psi$ mesons, as long as they are produced in the initial hard interaction.

Despite being studied for many decades~\cite{Jpsi1,Jpsi2,Jpsi3,Jpsi4,Jpsi5,Jpsi6,Jpsi7}, the production mechanism
of $J/\psi$ mesons in hadronic collisions is not fully understood. The main models for perturbative
calculations of heavy quarkonium production ($Q\bar{Q}$)
in hadronic collisions differ in whether the system is produced in a colour singlet (CS) state
or a colour octet (CO) state~\cite{Jpsi8,CASWELL1986437,Bodwin:1994jh,Mao:2013xpa,Chao:2012iv}.
The CS model requires two hard gluons in a colour singlet in the initial state, or one gluon splitting
into $Q\bar{Q}$ where one of the quarks radiates a hard gluon.
The non-relativistic QCD (NRQCD) framework allows the $Q\bar{Q}$ system to remain in a colour-octet state and then generates the final colour-neutral meson via low-energy non-perturbative matrix elements; these matrix elements are determined from fits to experimental data~\cite{CASWELL1986437,Bodwin:1994jh,Braaten:1999qk,Chao:2012iv,Butenschoen:2012qh}.
 
Associated prompt $J/\psi$ + $W^{\pm}$ production has been presented as a clear signature of CO
processes~\cite{PhysRevD.83.014001}, although other authors argue that higher-order CS processes
will dominate~\cite{bib:heavyQ}. The process $W^{\pm}\rightarrow W^{\pm}+\gamma^* \rightarrow J/\psi + W^{\pm}$
may contribute, but the focus for this measurement is a comparison to the CO processes~\cite{Lansberg:2013wva}.
The production rate measured by ATLAS at 7~\TeV, while having large statistical uncertainties, was an order of
magnitude larger than the NRQCD prediction of Ref.~\cite{PhysRevD.83.014001}.

This paper reports a measurement of the ratio of fiducial and inclusive cross-sections for associated
prompt $J/\psi$ + $W^\pm$ production to the cross-section of inclusive $W^\pm$ production in the same $W^\pm$
kinematic region.  The fiducial measurement for $J/\psi + W^\pm$ is defined in a restricted kinematic range
for the muons from $J/\psi$ decay, and is specific to the ATLAS detector, while the inclusive result is
determined by correcting for the detector's kinematic acceptance to muons. These cross-section ratios are
presented for $J/\psi$ transverse momenta in the range $8.5 < p_\mathrm{T} < 150$~\GeV\ and rapidities
satisfying $|y_{J/\psi}| < 2.1$.
The inclusive ratio is also quoted differentially as a function of the $J/\psi$
transverse momentum.

Single parton scattering (SPS) occurs in a given \textit{pp} collision when the $J/\psi$ meson and $W^\pm$
boson are produced from one parton pair, while
double parton scattering (DPS) occurs when the $J/\psi$ meson and $W^\pm$ boson are produced from two
different parton pairs.
The cross-section ratio for SPS is obtained after subtracting the estimated DPS fraction, and is
compared with a theoretical prediction of the next-to-leading-order CO contribution~\cite{Mao:2013xpa}.

\newcommand{\AtlasCoordFootnote}{
ATLAS uses a right-handed coordinate system with its origin at the nominal interaction point (IP)
in the centre of the detector and the $z$-axis along the beam pipe. The $x$-axis points from the IP
to the centre of the LHC ring, and the $y$-axis points upwards. Cylindrical coordinates $(r,\phi)$
are used in the transverse plane, $\phi$ being the azimuthal angle around the $z$-axis. The pseudorapidity
is defined in terms of the polar angle $\theta$ as $\eta = -\ln \tan(\theta/2)$. Angular distance is
measured in units of $\Delta R \equiv \sqrt{(\Delta\eta)^{2} + (\Delta\phi)^{2}}$.}

\section{ATLAS detector}
 
The ATLAS detector~\cite{PERF-2007-01} at the LHC is a multipurpose particle detector with a forward--backward
symmetric cylindrical geometry and a near $4\pi$ coverage in solid angle.\footnote{\AtlasCoordFootnote} It
consists of an inner tracking detector surrounded by a thin superconducting solenoid providing a \SI{2}{\tesla}
axial magnetic field, electromagnetic (EM) and hadron calorimeters, and a muon spectrometer. The inner tracking
detector covers the pseudorapidity range $|\eta| < 2.5$. It consists of silicon pixel, silicon microstrip, and
transition radiation tracking detectors. Lead/liquid-argon sampling calorimeters provide EM energy measurements
with high granularity. A steel/scintillator-tile hadron calorimeter covers the central pseudorapidity range $|\eta| < 1.7$.
The endcap and forward regions are instrumented with liquid-argon calorimeters for both EM and hadronic energy measurements
up to $|\eta| = 4.9$. The muon spectrometer surrounds the calorimeters and is based on three large air-core toroidal
superconducting magnets with eight coils each. The field integral of the toroids ranges between \num{2.0}
and \SI{6.0}{\tesla\metre} across most of the detector acceptance. The muon spectrometer includes a system
of precision tracking chambers and fast detectors for triggering.
 
A three-level trigger system was used to select events. The first-level trigger is implemented in hardware
and used a subset of the detector information to reduce the accepted rate to at most \SI{75}{\kilo\hertz}.
This was followed by two software-based trigger levels that together reduced the accepted event rate
to \SI{400}{\hertz} on average depending on the data-taking conditions during 2012~\cite{Aad:2012xs}.

\section{Event selection and reconstruction}
The analysis uses $20.3\,\mathrm{fb}^{-1}$ of $pp$ collision data at $\sqrt{s} = 8\,\mathrm{\TeV}$
collected during 2012. Events were selected
using a non-prescaled single-muon trigger that required at least one muon with |$\eta|<2.4$, transverse
momentum $p_\mathrm{T} > 24\,\mathrm{\GeV}$, stable beams, and fully operational subdetectors.
 
The muon reconstruction begins by finding a track candidate independently in the inner tracking detector
and the muon spectrometer. The momentum of the muon candidate is calculated by statistically combining the
information from the two subsystems and correcting for parameterised energy loss in the calorimeter;
these muon candidates are referred to as \textit{combined} muons.
 
In some cases a track in the inner detector is identified as a muon if the extrapolated track is associated
with at least one local track segment in the muon spectrometer.  In such cases the information from the inner
tracking detector alone is used to determine the momentum.
For analyses studying low-mass objects, such as $J/\psi$ mesons, the inclusion of these \textit{segment-tagged}
muons provides additional efficiency for reconstructing low-$p_\mathrm{T}$ muons~\cite{PERF-2014-05}.

\subsection{$W^\pm$ selection}
\renewcommand\arraystretch{1.2}
\begin{table}[h!]
\begin{center}
\caption[]{Selection criteria for the inclusive $W^{\pm}$ sample, where $\mu$ is the muon from the $W^\pm$ boson decay.}
\begin{tabular}{rl}
\hline
\multicolumn{2}{c}{$W^\pm$ boson selection}									\\
\hline
\multicolumn{2}{c}{At least one isolated muon that originates $< 1$ mm from primary vertex along $z$-axis}
\\
\multicolumn{2}{c}{$p_\mathrm{T}$ (trigger muon) $> 25\,\mathrm{\GeV}$} 					\\
\multicolumn{2}{c}{$|\eta^\mu|< 2.4$} 									\\
\multicolumn{2}{c}{Missing transverse momentum $ > 20\, \mathrm{\GeV}$}            			\\
\multicolumn{2}{c}{$m_\mathrm{T}(W^\pm) >  40\, \mathrm{\GeV}$}						\\
\multicolumn{2}{c}{$|d_0|/\sigma_{d_0} < 3$} 					\\
\hline
 
\end{tabular}
\label{tab:WTab}
\end{center}
\end{table}
 
\renewcommand\arraystretch{1.2}
\begin{table}[h!]
\begin{center}
\caption[]{Definition of the fiducial region for the $J/\psi$ cross section measurement, where $\mu_1$ is the highest-$p_\mathrm{T}$ muon from the $J/\psi$ decay,
and $\mu_2$ is the second-highest-$p_\mathrm{T}$ muon from the $J/\psi$ decay.}
\begin{tabular}{rl}
\hline
\multicolumn{2}{c}{$J/\psi$ selection}												\\
\hline
\multicolumn{2}{c}{$ 2.4 < m(\mu^+\mu^-) < 3.8 $ \GeV} 										\\
\multicolumn{2}{c}{$ 8.5 < p_\mathrm{T}^{J/\psi} < 150 $ \GeV,  $|y_{J/\psi}|<2.1$}						\\
\multicolumn{2}{c}{$p_\mathrm{T}^{\mu_1} > 4 $ \GeV, $|\eta^{\mu_1}|< 2.5$}                                                         \\
\ldelim\{{2}{3mm}[] either $p_\mathrm{T}^{\mu_2}> 2.5$ \GeV, & 1.3 $ \leq |\eta^{\mu_2}|< 2.5$ \rdelim\}{2}{3mm}[]		\\
or $p_\mathrm{T}^{\mu_2}> 3.5$ \GeV, &  $|\eta^{\mu_2}|< 1.3$						\\
\hline
\end{tabular}
\label{tab:jpsiTab}
\end{center}
\end{table}
 
\renewcommand\arraystretch{1.0}
An inclusive $W^\pm$ sample is defined by applying the $W^\pm$ boson selections listed in Table \ref{tab:WTab}.
Candidate muons from $W^\pm$ decays are required to be \textit{combined} and to match the muon reconstructed by the
trigger algorithm. The primary vertex is chosen as the reconstructed vertex with the highest $\Sigma p_\mathrm{T}^2$ of
associated tracks and must have a minimum of three associated tracks with $p_\mathrm{T} > 400\,\mathrm{\MeV}$.
 
Calorimetric and track isolation variables are defined by calculating the sum of transverse energy $(E_\mathrm{T})$
deposits in the calorimeter cells and track $p_\mathrm{T}$, respectively, within a cone size $\Delta R = 0.3$ around the
muon direction. The energy deposited by the muon is subtracted from the calorimetric isolation variable, and only
tracks compatible with originating from the primary vertex and with $p_\mathrm{T} >$ 1~\GeV\ (excluding the muon itself)
are considered for the track isolation. A correction depending on the number of reconstructed vertices is made to the
calorimetric isolation to account for additional energy deposits due to pile-up vertices.\footnote{Pile-up arises from multiple proton--proton
collisions that occur in the same bunch crossing.}  For the muon to be considered isolated, the two isolation variables
defined above must both be less than 5\% of the muon $p_\mathrm{T}$.
 
Transverse impact parameter significance is defined as $|d_0|/\sigma_{d_0}$, where $d_0$ is the impact parameter, defined
as the distance of closest approach of the muon trajectory to the primary vertex in the $xy$-plane,
and $\sigma_{d_0}$ is its uncertainty.
 
The $W^\pm$ boson transverse mass is defined as
\begin{equation*}
m_\mathrm{T}(W^\pm)\equiv\sqrt{2p_\mathrm{T}(\mu)E^\mathrm{miss}_\mathrm{T}[1-\cos(\phi^\mu-\phi^\nu)]}\text{ , }
\end{equation*}
where the variables $\phi^\mu$ and $\phi^\nu$ represent the azimuthal angles of the muon from the $W^\pm$ boson decay and
the missing transverse momentum $E_\mathrm{T}^\mathrm{miss}$, respectively.
The $E_\mathrm{T}^\mathrm{miss}$ is calculated as the magnitude of the negative vector sum of the transverse momenta of
calibrated electrons, photons, hadronically decaying $\tau$-leptons, jets and muons, as well as additional low-momentum
tracks that are associated with the primary vertex but are not associated with any other $E_\mathrm{T}^\mathrm{miss}$ component~\cite{Aad:2016nrq}.

\subsection{$W^\pm$ + $J/\psi$ event selection}
If an event has two additional muons then the $J/\psi$ selections listed
in Table \ref{tab:jpsiTab} are also applied to define the
associated $J/\psi$ + $W^\pm$ sample.
The $J/\psi$ candidates are required to have a vertex $< 10$ mm from the primary vertex along the $z$-axis and  must be
formed from either two \textit{combined} muons or from one \textit{combined} muon and one \textit{segment-tagged} muon,
and at least one muon must have $p_\mathrm{T} >4\,\mathrm{\GeV}$. A vertex fit is performed to constrain the two muons to originate from a common point.
 
To distinguish prompt $J/\psi$ candidates from those originating from $b$-hadron decay (non-prompt), the pseudo proper decay time is used:

\begin{equation*}
\tau(\mu^+\mu^-)\equiv\frac{\vec{L} \cdot \vec{p}_\mathrm{T}^{\;J/\psi}}{p_\mathrm{T}^{J/\psi}} \cdot \frac{m(\mu^+\mu^-)}{p_\mathrm{T}^{J/\psi}}\text{ , }
\end{equation*}
 
where $\vec{L}$ is the 2-D displacement vector of the $J/\psi$ decay vertex from the primary event vertex, and $\vec{p}_\mathrm{T}^{\;J/\psi}$ and $m(\mu^+\mu^-)$
are the transverse momentum and invariant mass of the $J/\psi$ candidate, respectively. Prompt $J/\psi$ candidates should have a pseudo proper
decay time consistent with zero (within resolution).

\section{Signal and background extraction}
 
\subsection{Inclusive $W^{\pm}$  sample}
 
A signal sample of $W^\pm \rightarrow \mu \nu$ Monte Carlo (MC) was used to verify the overall modelling of the signal+background in the inclusive $W^\pm$ sample.
The backgrounds $W^\pm \rightarrow\tau\nu$, $Z\rightarrow\mu\mu$, $Z\rightarrow\tau\tau$, diboson, $t\bar{t}$ and single top were also
modelled with MC simulations.
Most of the MC samples were generated using \POWHEGBOX~\cite{powheg1,powheg2,powheg3} for the hard scatter and showered using
either \PYTHIAV{6}~\cite{pythia6} or \PYTHIAV{8}~\cite{bib:pythia}.
Samples of $W$ or $Z$ bosons decaying into electrons, muons or taus were generated with the \POWHEGBOX next-to-leading-order (NLO)
generator, interfaced
to \PYTHIAV{8} with the AU2 set of tuned parameters~\cite{ATL-PHYS-PUB-2012-003} for the underlying event
and the CT10 leading-order (LO) parton distribution function (PDF) set~\cite{Lai:2010vv}.  Processes involving $t\bar{t}$
and single top were generated with \POWHEGBOX\ using the CT10 PDFs, interfaced to \PYTHIAV{6.427} with the P2011C underlying-event
tune~\cite{Skands:2010ak} and the CTEQ6L1 PDF set~\cite{Pumplin:2002vw}.  Diboson samples were produced with \HERWIGV{6.520.2}~\cite{bib:herwig}
with the ATLAS AUET2 underlying-event tune~\cite{ATL-PHYS-PUB-2011-008} and CTEQ6L1.
Alternative samples are used to evaluate the systematic uncertainties: \ALPGENV{2.13}~\cite{bib:alpgen} with \HERWIGV{6.520.2}
parton showering with CTEQ6L1 for $W$+jets and $Z$+jets, including \JIMMY~\cite{ATL-PHYS-PUB-2010-014} for multiparton
interactions, \MCatNLOV{4.06}~\cite{bib:mcatnlo} with \HERWIGV{6.520} parton showering for $t\bar{t}$, and \ACERMC~\cite{bib:acermc}
with \PYTHIAV{6.426}~\cite{pythia6} and CTEQ6L1 for single top.
All simulated samples were processed through a \textsc{Geant4}-based detector simulation~\cite{Agostinelli:2002hh,SOFT-2010-01}
with the standard ATLAS reconstruction software used for collision data.

For the multijet background, a standard data-driven technique called the $ABCD$ method~\cite{BPHY-2012-06} is used.
Four independent regions ($A$, $B$, $C$, $D$)
are defined in a two-dimensional plane using  $m_\mathrm{T}(W^\pm)$ and $E_\mathrm{T}^\mathrm{miss}$ together with
the uncorrelated muon isolation variable.
Regions $A$ and $B$ are required to have $E_\mathrm{T}^\mathrm{miss} < $ 20~\GeV\ and  $m_\mathrm{T}(W^\pm) < $ 40~\GeV,
while regions $C$ and $D$ are required to have $E_\mathrm{T}^\mathrm{miss} >$ 20~\GeV\ and $m_\mathrm{T}(W^\pm) >$ 40~\GeV.
In regions $A$ and $C$ ($B$ and $D$) an isolated muon
(non-isolated muon) is required. The multijet background in signal region $C$ is determined
from $N_C = N_A \times N_D/N_B$, where $N_A$, $N_B$, $N_C$, and $N_D$ are the background-subtracted event yields in
regions $A$, $B$, $C$ and $D$ respectively.

After accounting for all background events (which contribute an estimated 12\% of the original yield, with $Z\rightarrow \mu^+ \mu^-$
and $W^{\pm}\rightarrow \tau^{\pm} \nu$ making up 80\% of the background), a total $W^\pm$ yield of $(6.446 \pm 0.035) \times 10^7$ events is found.  The uncertainty includes the statistical uncertainty in the data sample and systematic uncertainties arising from the background sample sizes, background cross-sections, the multijet estimation and the luminosity uncertainty.
The absolute luminosity scale is derived from beam-separation scans performed in November 2012. The uncertainty in the integrated luminosity is 1.9\%~\cite{DAPR-2013-01}.

\subsection{Separation of prompt and non-prompt $J/\psi$}
The associated prompt $J/\psi$ + $W^{\pm}$ yield is measured using a two-dimensional unbinned maximum likelihood fit to the $J/\psi$ mass and pseudo proper decay time
in the region 2.4~\GeV\ $<$ $m(\mu^+\mu^-)$ $<$ 3.8~\GeV\ and
$-2$ ps $<$ $\tau(\mu^+\mu^-)$ $<$ $10$ ps.
The pseudo proper decay time for the prompt signal is modelled as a double Gaussian distribution while a single-sided exponential function is used for
the non-prompt signal. The prompt background component is modelled as a double-sided exponential function and the non-prompt background is the sum of a
single-sided and a double-sided exponential function. The lifetime fit takes into account resolution effects by convolving the exponential functions with a
Gaussian resolution function.  The $J/\psi$ mass distribution is modelled with a Gaussian distribution for both the prompt and non-prompt signal and a
third-order polynomial is used for both the prompt and non-prompt combinatorial backgrounds.  To improve the stability of the fit, the mean and width
of the $J/\psi$ mass distribution are fixed to the values derived from fitting a large inclusive $J/\psi$ sample.
 
After the fit is performed, the sPlot tool~\cite{Pivk:2004ty} is used to extract per-event weights according to the parameters of the fit model.
These weights are used to generate prompt signal distributions for other variables such as the $W^\pm$ transverse mass, the $J/\psi$ transverse
momentum and the azimuthal opening angle between the $W^\pm$ and the $J/\psi$.
 
\begin{figure}[h!]
\centering
\subfloat[]{
\includegraphics[width=0.45\textwidth]{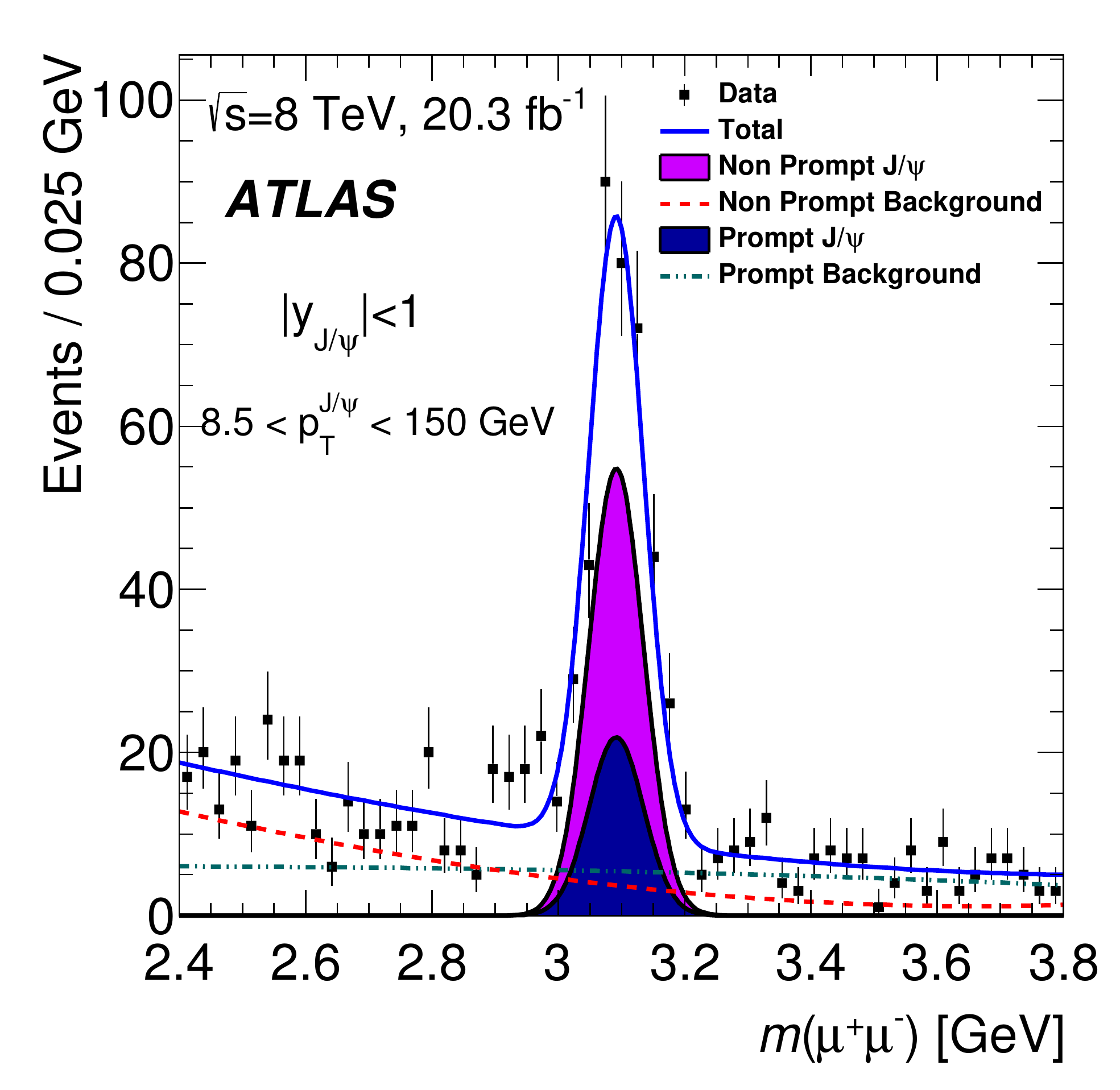}
}
\subfloat[]{
\includegraphics[width= 0.45\textwidth]{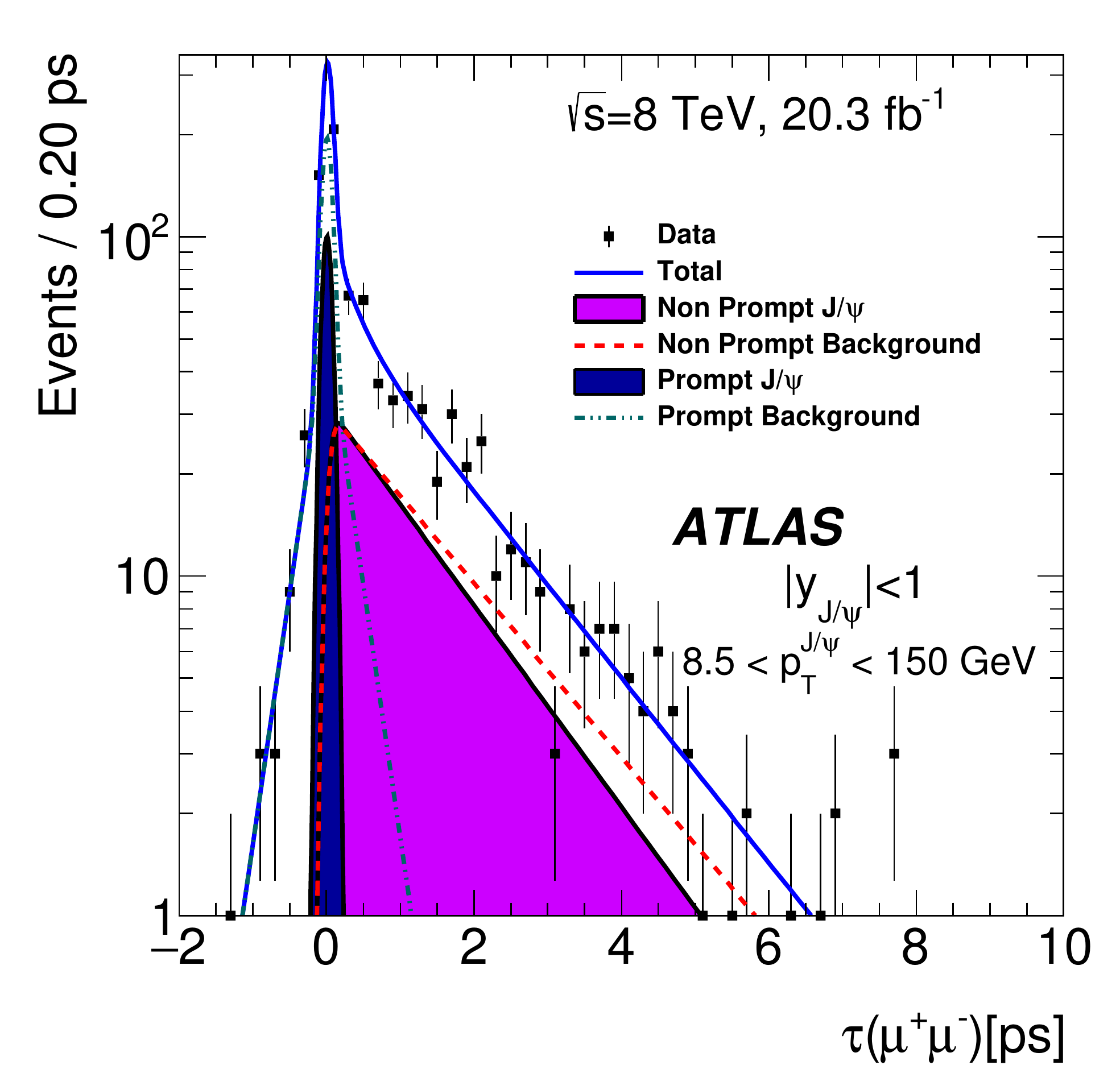}
}
 
\subfloat[]{
\includegraphics[width=0.45\textwidth]{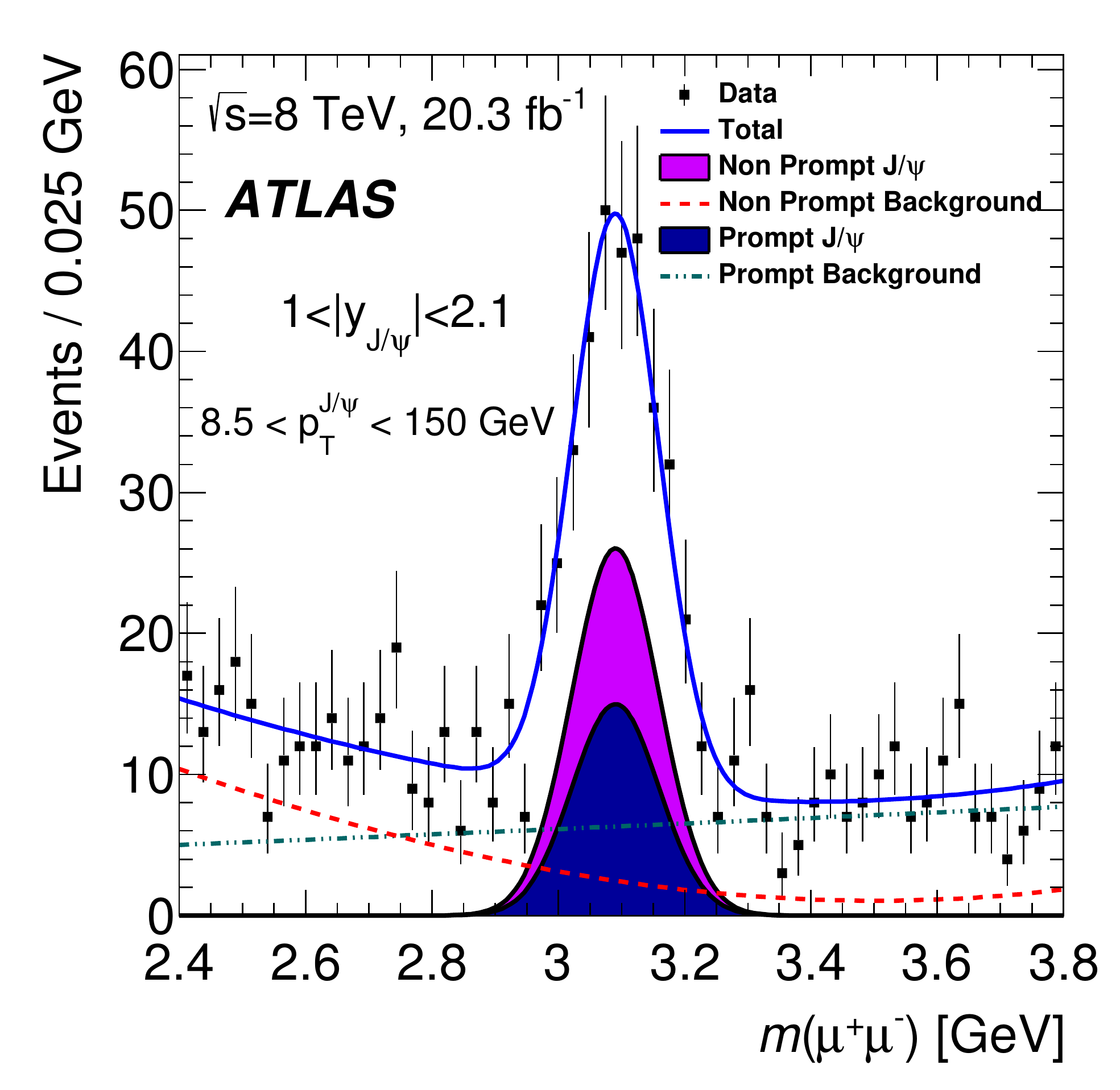}
}
\subfloat[]{
\includegraphics[width= 0.45\textwidth]{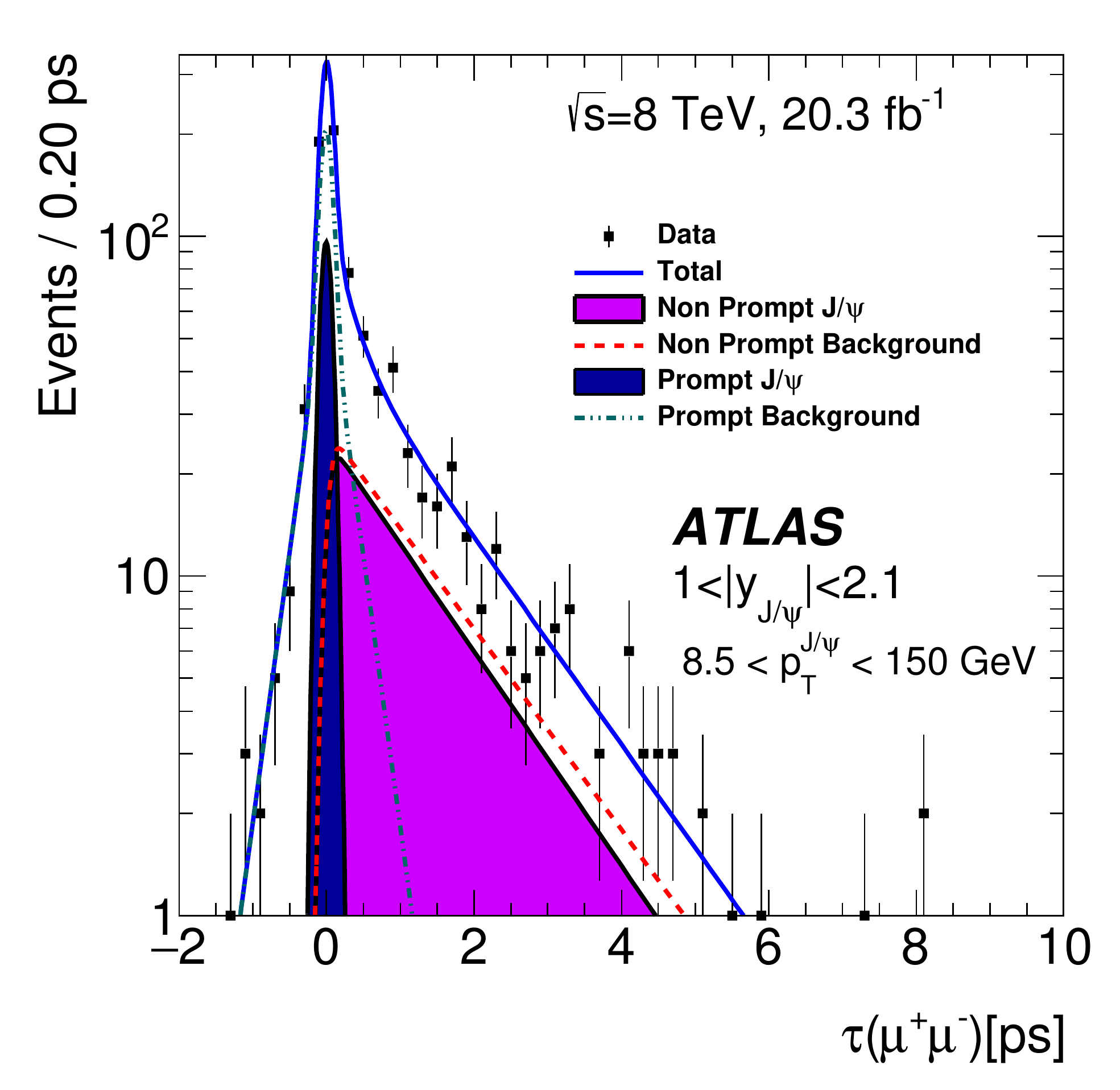}
}
\caption{(a) $J/\psi$ candidate mass and (b) pseudo proper decay time for the rapidity range $|y_{J/\psi}|<1$ and
$p_T$ range $8.5 < p_\mathrm{T}^{J/\psi}< 150$ GeV;
(c) $J/\psi$ candidate mass and (d) pseudo proper decay time for the rapidity range $1<|y_{J/\psi}|<2.1$
and  $p_T$ range $8.5 < p_\mathrm{T}^{J/\psi}< 150$ GeV.}
\label{fig:muon_tau2}
\end{figure}

The results of applying the two-dimensional mass and lifetime fit to the $J/\psi$ candidate events are shown in Figure~\ref{fig:muon_tau2}, giving prompt signal yields of  93 $\pm$ 14 (stat) for $|y_{J/\psi}|<1$ and 102 $\pm$ 17 (stat) for $1<|y_{J/\psi}|<2.1$.
Two rapidity ranges are used to account for the difference in muon momentum resolution between the barrel and endcap regions of the detector.

\subsection{$W^\pm + J/\psi$ backgrounds}
The same backgrounds considered for the inclusive $W^{\pm}$ sample are used
for the associated prompt $J/\psi + W^\pm$ sample.  In addition, background from
$B_c \rightarrow J/\psi \mu \nu$ is also considered.
Using MC, the expected
yields are found to be consistent with zero ($3.7^{+1.9}_{-3.4}$ events).
A significant background arises from simultaneous production of a $W^{\pm}$
and a $J/\psi$ from different $pp$ interactions in the same bunch crossing,
where the two production vertices are not distinguished.
The probability that, when a $W^{\pm}$ is
produced, a $J/\psi$ is also produced nearby, can be estimated statistically.
The average number of pile-up collisions occurring within 10 mm of a given
interaction vertex is determined to be 2.3 $\pm$ 0.2 and is found by sampling the luminosity-weighted
distribution of the mean number of inelastic interactions per proton--proton
bunch crossing.  This number is combined with the $pp$ inelastic cross-section and the
prompt $J/\psi$ cross-section~\cite{BPHY-2014-01} to give
an estimate of
the pile-up contribution as a function of the $p_{\mathrm{T}}$ and rapidity of the $J/\psi$ in the
associated production sample.
The fraction of pile-up events is determined to be (10.5 $\pm$ 1.2)\% of the candidate events.
 
The desired signal topology is prompt $J/\psi+W^{\pm}$, where the
$W^{\pm}$ boson decays to $\mu^{\pm}\nu$.  Production of prompt
$J/\psi+W^{\pm}$ with a different decay of the $W^{\pm}$ boson, or
of prompt $J/\psi+Z$, are treated as backgrounds.
Background from prompt $J/\psi+W^\pm$ with $W^\pm\rightarrow \tau^\pm \nu$  is determined using MC.
An inclusive MC sample of $W^\pm\rightarrow \tau^\pm \nu$  events is used to determine the probability of an event
to pass the $W^\pm \rightarrow \mu^\pm \nu$ selection, yielding a background of (2.3 $\pm$ 0.1)\%
of the candidate events.
Background from prompt $J/\psi+Z$ events is calculated using the measured value of
$\sigma(pp \rightarrow J/\psi+Z$)/$\sigma(pp \rightarrow Z)$ in the 8 TeV ATLAS data~\cite{BPHY-2014-01}.
This ratio is scaled by the probability of  $Z\rightarrow \mu^+ \mu^-$ and $Z\rightarrow \tau^+ \tau^-$
to pass the  $W^\pm \rightarrow \mu^\pm \nu$ selection in inclusive MC samples, giving a total background
of (9.5 $\pm$ 0.5)\% events.
The $J/\psi+Z$ background is subtracted as a constant fraction in the $p_{\mathrm{T}}$ differential
distribution since the measured
ratio between $\sigma(pp\rightarrow J/\psi+Z)/\sigma(pp \rightarrow Z)$ and $\sigma(pp\rightarrow J/\psi+W^{\pm})/
\sigma(pp\rightarrow W^{\pm}$) is consistent
with being flat as a function of $p_\mathrm{T}^{J/\psi}$.
 
\subsection{Detector effects and acceptance corrections}
 
The efficiency for reconstructing muons varies depending on the $p_{\mathrm{T}}$ of the muon, with efficiencies of 65\% for 3~\GeV\
muons increasing to a plateau efficiency of 99\% for muons above 10~\GeV.  The nominal relative momentum resolution for muons is $<3.5\%$
up to transverse momenta $p_\mathrm{T}\sim200$~\GeV~\cite{PERF-2011-01}. To correct the measurements for reconstruction efficiency,
a per-event weight is computed using muon efficiency measurements extracted from large inclusive $J/\psi\rightarrow\mu^+\mu^-$
and $Z\rightarrow\mu^+\mu^-$ data samples and applied as a function of the pseudorapidity and $p_{\mathrm{T}}$ of each muon from the $J/\psi$ decay \cite{BPHY-2014-01}.
In addition, a per-event weight is applied to correct the $J/\psi$ rate for muons that fall outside the detector acceptance.
The acceptance weight is given by the probability that both muons in a $J/\psi\rightarrow\mu^+\mu^-$ candidate pass the kinematic requirements on $p_\mathrm{T}^{\mu}$ and $|\eta^{\mu}|$, for a particular $y_{J/\psi}$ and $p_\mathrm{T}^{J/\psi}$.  These weights are determined using generator-level simulations.
Although inclusive $J/\psi$ spin-alignment measurements find a near isotropic distribution~\cite{CMS-BPH-13-003,Abelev:2011md,Aaij:2013nlm},
this may not apply to the spin-alignment of $J/\psi$ mesons produced in association with a $W$ boson, due to the different
relative contributions of the $J/\psi$ production modes.  Consequently, a nominal uniform spin-alignment is used and a
variety of extreme polarisation states of the $J/\psi$ are considered for
the acceptance correction, one with full longitudinal polarisation and three with different transverse polarisations~\cite{BPHY-2014-01}.

After correcting for the  $J/\psi$ daughter muon efficiency and acceptance, ratios of
cross-sections for associated prompt $J/\psi$ + $W^{\pm}$ production to
inclusive $W^{\pm}$ production are measured in a single
$W^{\pm}\rightarrow \mu^{\pm} \nu$ fiducial region defined as $|\eta^{\mu}|$ < 2.4, $p_\mathrm{T}(\mu^{\pm})$ > 25 GeV and $p_\mathrm{T}(\nu)$ > 20 GeV, both differentially in $p_\mathrm{T}^{J/\psi}$ and
also integrated over $p_\mathrm{T}^{J/\psi}$.  These measurements will be discussed in
Section 6.
Using MC, the efficiency for reconstructing inclusive
$W^{\pm} \rightarrow \mu \nu$ is found to depend linearly on the
$p_{\mathrm{T}}$ of the $W^{\pm}$ boson ($p_\mathrm{T}^W$).
A linear correlation is also found between the values of $p_\mathrm{T}^{J/\psi}$ and $p_{\mathrm{T}}^W$ for the associated
production sample in data.  These two effects
lead to a correction to the differential cross-section ratio based on the
$p_{\mathrm{T}}$ of the prompt $J/\psi$ candidate.
To apply the correction, the average value of $p_\mathrm{T}^{J/\psi}$ is determined for each $p_\mathrm{T}^{J/\psi}$ bin in the differential distribution. The linear correlation between $p_\mathrm{T}^{J/\psi}$ and $p_{\mathrm{T}}^W$ is used to derive the corresponding value for the average $p_{\mathrm{T}}^W$ within the $p_\mathrm{T}^{J/\psi}$ bin.
The ratio of the inclusive $W^{\pm}$ efficiency to the $W^{\pm}$ reconstruction efficiency in each $p_\mathrm{T}^{J/\psi}$ bin gives the efficiency correction, which varies
from $0.93 \pm 0.02$ at  low $p_\mathrm{T}^{J/\psi}$ to $0.78 \pm 0.04$ in the highest  $p_\mathrm{T}^{J/\psi}$ bin.

\subsection{Double parton scattering}
 
The measured yield of  prompt $J/\psi$ + $W^{\pm}$ includes contributions from SPS and DPS processes. The DPS contribution
can be estimated using the effective cross-section ($\sigma_{\mathrm{eff}}$) measured by the ATLAS Collaboration,
as well as the double-differential cross-section for $pp\rightarrow J/\psi$ prompt production
($\sigma_{J/\psi}$)~\cite{BPHY-2014-01}. Based on the assumptions that the two hard scatters are uncorrelated,
the probability that a $J/\psi$ is produced by a second hard process in an
event containing a $W^{\pm}$ boson is given by

\begin{equation*}
P^{ij}_{J/\psi|W^{\pm}}=\frac{\sigma^{ij}_{J/\psi}}{\sigma_{\mathrm{eff}}}\text{ , }
\end{equation*}
 
where $\sigma^{ij}_{J/\psi}$ is the cross-section for ${J/\psi}$ production in the appropriate $p_\mathrm{T}$ ($i$) and rapidity ($j$) interval and $\sigma_{\mathrm{eff}}$
is the effective transverse overlap area of the interacting partons.
Since $\sigma_{\mathrm{eff}}$ may not be process-independent, it is unclear which
value of $\sigma_{\mathrm{eff}}$ to use for prompt $J/\psi$ + $W^{\pm}$ production, so two different values are considered:
$\sigma_{\mathrm{eff}} = 15 \pm 3 (\mathrm{stat.}) ^{+5}_{-3} (\mathrm{sys.})$~mb from
$W^\pm+2$-jet events~\cite{STDM-2011-29}
and $\sigma_{\mathrm{eff}} = 6.3 \pm 1.6 (\mathrm{stat.}) \pm 1.0 (\mathrm{sys.})$~mb from prompt $J/\psi$ pair production~\cite{BPHY-2015-02}.
These two values of $\sigma_{\mathrm{eff}}$ are chosen since they are the two ATLAS measurements closest to the $J/\psi + W^{\pm}$ final state.
The latter value is close to those inferred in Refs.~\cite{Lansberg:2016rcx,Lansberg:2017chq} from the earlier ATLAS measurements
of $J/\psi$ + $W^{\pm}$ and $J/\psi$ + $Z$ production~\cite{BPHY-2012-06,BPHY-2014-01}. With these assumptions, it is estimated that
between $(31^{+9}_{-12})\%$ ($\sigma_{\mathrm{eff}}$ = 15 mb) and $(75 \pm 23)\%$ ($\sigma_{\mathrm{eff}}$ = 6.3 mb) of the inclusive
signal yield is due to DPS interactions, where the uncertainties in the inclusive $W^{\pm}$ yield, the $J/\psi$ cross-section
and $\sigma_{\mathrm{eff}}$ are propagated to the DPS fraction.
 
\begin{figure}[h]
\centering
\includegraphics[width=0.45\textwidth]{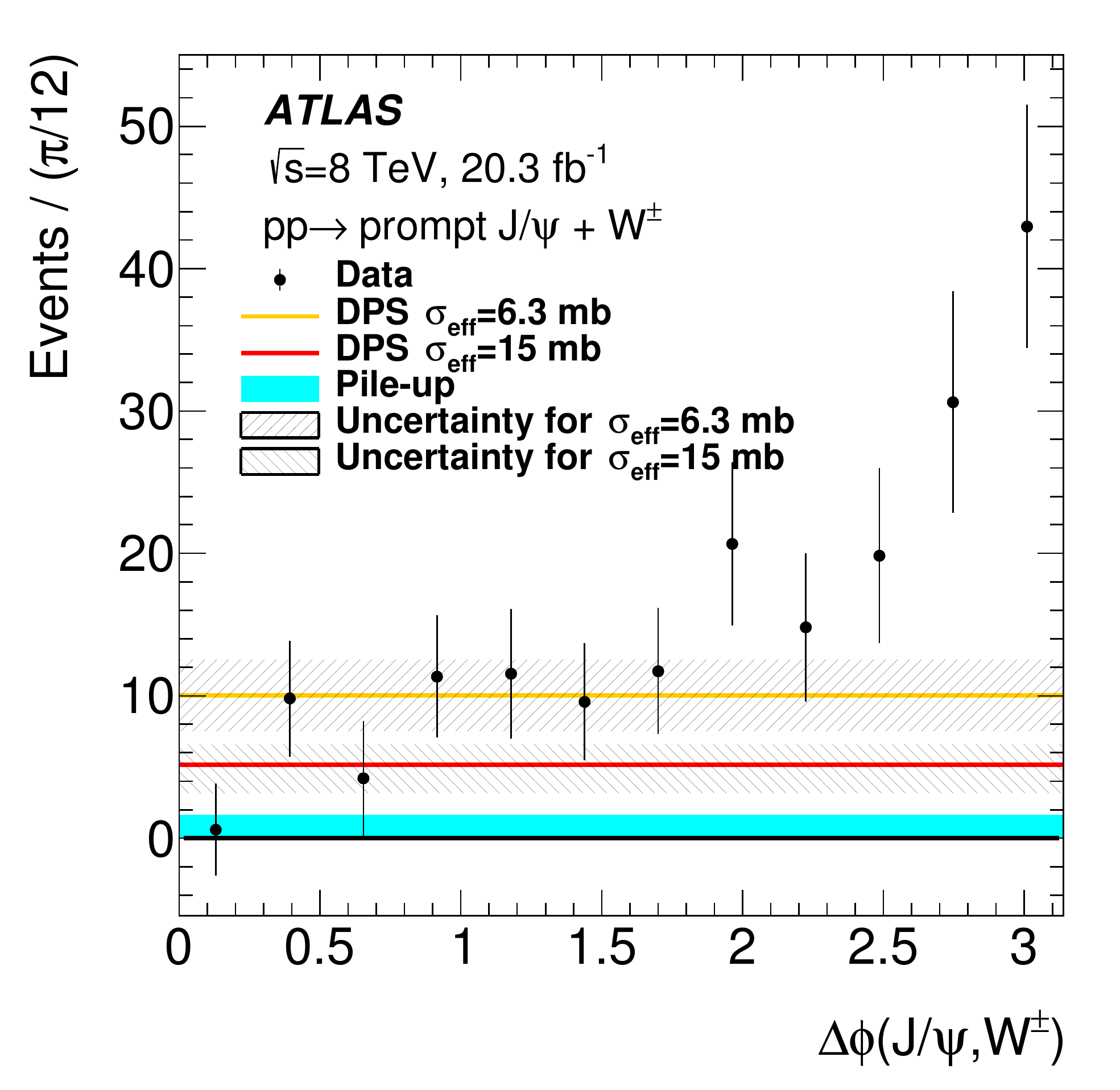}
\caption{The sPlot-weighted opening angle $\Delta\phi(J/\psi,W^\pm)$ for prompt $J/\psi+W^\pm$ candidates,
uncorrected for efficiency or acceptance, compared with the sum of the expected pileup and DPS contributions.  The data are not corrected for
$J/\psi+V$ backgrounds which contribute $\sim$10\% and have a shape similar to the overall distribution.
The DPS contribution is shown for two $\sigma_{\mathrm{eff}}$ values, $15$~mb and $6.3$~mb, as described in the text.
The peak at $\Delta\phi\simeq\pi$ is assumed to come primarily from SPS events.}
\label{fig:DeltaPhisPlot}
\end{figure}

The distribution of the azimuthal opening angle $\Delta\phi(J/\psi,W^\pm)$ between the directions of the $J/\psi$ and of the $W^\pm$
is sensitive to the contributions of SPS and DPS\@. The DPS component should not have a preferred $\Delta\phi$  value,
while the SPS events are expected to peak at $\Delta\phi\approx\pi$ due to momentum conservation.
The estimated DPS yield can be validated with data, assuming that the low $\Delta\phi(J/\psi,W^\pm)$ is exclusively
due to DPS interactions.
Figure~\ref{fig:DeltaPhisPlot} shows the measured $\Delta\phi$ distribution with the estimated DPS contribution using the two
different values of $\sigma_{\mathrm{eff}}$.
Both values of $\sigma_{\mathrm{eff}}$ are consistent with the data at low $\Delta\phi$.   The normalized  $\Delta\phi$
distributions with and without correcting for efficiency and acceptance are consistent with each other within the statistical uncertainties
 
\section{Systematic uncertainties}
 
Almost all systematic uncertainties associated with the reconstruction of the $W^\pm$ boson and the integrated luminosity cancel out in
the ratio of the two processes, $J/\psi$ + $W^{\pm}$ and inclusive $W^\pm$ production, in the same fiducial region. The remaining relevant
systematic uncertainties are discussed below.
 
The choice of functions used to fit the mass and pseudo proper decay time is a source of systematic uncertainty.
Three alternative models for the mass fit are studied: introducing a $\psi(2S)$ mass peak into the fit model, letting the mean of the $J/\psi$
mass peak float, and using exponential functions to model the background. The maximum difference between the nominal model yield and the yields
from the alternative fit models is taken as a systematic uncertainty. An alternative pseudo proper decay time model which takes into account
resolution effects by convolving the lifetime with a double Gaussian resolution function was found not
to make a significant difference to the prompt $J/\psi$ yield.
 
The reconstruction efficiencies used for the muons from $J/\psi$ decay are derived from data as a function of $p_\mathrm{T}$ and $\eta$ as
discussed in the previous section. A systematic uncertainty is determined by randomly varying the efficiency in each $p_\mathrm{T}$--$\eta$
interval 100 times using a Gaussian distribution of width equal to the uncertainty in the efficiency in that interval. The RMS spread of
the extracted yield is taken as the systematic uncertainty. The uncertainty due to the pile-up background estimation is also considered.
 
The $J/\psi$ vertex is required to be within 10 mm of the primary vertex along the $z$-axis, which can affect the pseudo proper decay time
distribution.
The impact of this is determined by taking the difference in yields between the nominal value of 10 mm and a value of 20 mm, after correcting for pileup contributions, and included as a systematic uncertainty.

The uncertainty on the fractional background from prompt $J/\psi+W^{\pm}$ with $W\rightarrow \tau^{\pm} \nu$
is determined by propagating the statistical and systematic uncertainties on the numbers of selected $W^{\pm}\rightarrow \tau^{\pm}\nu$ and
$W^{\pm}\rightarrow \mu^{\pm}\nu$
events in the inclusive MC samples.  The background correction for prompt $J/\psi+Z$ contamination incorporates the uncertainties on
the selected
$Z\rightarrow \mu^+ \mu^-$, $Z\rightarrow \tau^+\tau^-$ and $W^{\pm}\rightarrow \mu^{\pm}\nu$ events in the same way, and combines this
with the full uncertainty (statistical and systematic) from the $\sigma(pp\rightarrow \mathrm{prompt}\; J/\psi+Z)/\sigma(pp \rightarrow Z)$
measurement~\cite{BPHY-2014-01}.

The uncertainty on the difference in the reconstruction efficiency between the
inclusive $W^{\pm}$ sample and the prompt $J/\psi + W^{\pm}$ sample takes
several effects into account: the spread of $p_\mathrm{T}^{J/\psi}$ in each bin of the differential distribution;
the uncertainties in the linear fit for the reconstruction efficiency as a function of $p_\mathrm{T}^W$; and the uncertainties in the fit to determine $p_\mathrm{T}^W$ as a function of $p_\mathrm{T}^{J/\psi}$.
 
A nominal uniform spin-alignment is used; however, five different spin-alignment scenarios are considered, following the procedure
adopted and described in detail in Ref.~\cite{BPHY-2014-01},
leading to a systematic uncertainty due to the unknown spin-alignment.
A summary of the systematic uncertainties is given in Table~\ref{tab:syst}.
The effects of the different spin-alignment assumptions are shown in Tables~\ref{tab:acceptance_corrections_central}-\ref{tab:acceptance_corrections}.
 
\begin{table}[htb]
\caption{Summary of the systematic uncertainties, expressed as a percentage of the measured inclusive cross-section ratio of $J/\psi$ + $W^{\pm}$ to $W^{\pm}$.}
\begin{center}
\begin{tabular}{lSS}
\hline
Source of Uncertainty		& \multicolumn{2}{c}{Uncertainty [\%]} \\
& {$|y_{J/\psi}|<1$} &  {$1<|y_{J/\psi}|<2.1$} \\
\hline
$J/\psi$ mass fit  			&  8.7		& 4.9	\\
Vertex separation			&  12		        & 15 	\\
$\mu_{J/\psi}$ efficiency 		&  2.0		& 1.6	\\
Pile-up					&  1.1		& 1.4 \\
$J/\psi+Z$ and $J/\psi+W^\pm(\rightarrow \tau^\pm  \nu)$ & 3.5 & 4.8 \\
Efficiency correction                   &  2.3          & 2.3 \\
 
\hline

\end{tabular}
\end{center}
\label{tab:syst}
\end{table}

\begin{table*}[htbp]
\caption{
Percentage variations on the differential distribution for four extreme cases of
$J/\psi$ spin alignment of maximal polarisation relative to the
nominal unpolarised assumption for $|y_{J/\psi}|<1$~\cite{BPHY-2014-01}.}
\begin{center}
\begin{tabular}{ c c c c c}
\toprule
$p_T^{J/\psi}$ range [GeV] & Longitudinal &  Transverse~$0$ & Transverse~$+$ &Transverse~$-$ \\
\midrule
{$\begin{aligned}[t]
(8.5&,10) \\
(10&,14) \\
(14&,18)\\
(18&,30) \\
(30&,60) \\
(60&,150) \\
\end{aligned}$}
&
{$\begin{aligned}[t]
11&      \\
8&.9     \\
12&      \\
8&.1     \\
2&.3     \\
5&.2     \\
\end{aligned}$}
&
{$\begin{aligned}[t]
-4&.4  \\
-3&.1  \\
-5&.0  \\
-3&.3  \\
-0&.7  \\
-2&.2  \\
\end{aligned}$}
&
{$\begin{aligned}[t]
40&     \\
33&     \\
24&     \\
18&     \\
11&     \\
4&.0    \\
\end{aligned}$}
&
{$\begin{aligned}[t]
-28&      \\
-25&      \\
-23&      \\
-18&      \\
-10&      \\
-8&.0     \\
\end{aligned}$}
\\
\midrule
Total  & \hspace*{1mm} $9.6$ & $-3.7$ & $31$ & $-24$ \\
\bottomrule
\end{tabular}
\end{center}
\label{tab:acceptance_corrections_central}
\end{table*}

\begin{table*}[htbp]
\caption{
Percentage variations on the differential distribution for four extreme cases of
$J/\psi$ spin alignment of maximal polarisation relative to the
nominal unpolarised assumption for $1< |y_{J/\psi}|<2.1$~\cite{BPHY-2014-01}.}
\begin{center}
\begin{tabular}{ c c c c c}
\toprule
$p_T^{J/\psi}$ range [GeV] & Longitudinal &  Transverse~$0$ & Transverse~$+$ &Transverse~$-$ \\
\midrule
{$\begin{aligned}[t]
(8.5&,10) \\
(10&,14) \\
(14&,18)\\
(18&,30) \\
(30&,60) \\
(60&,150) \\
\end{aligned}$}
&
{$\begin{aligned}[t]
-19&     \\
-19&     \\
-15&     \\
-13&     \\
-7&.9    \\
-4&.8    \\
\end{aligned}$}
&
{$\begin{aligned}[t]
13&      \\
12&      \\
9&.8     \\
8&.0     \\
4&.6     \\
2&.6    \\
\end{aligned}$}
&
{$\begin{aligned}[t]
38&     \\
28&     \\
18&     \\
11&     \\
7&.4    \\
3&.9    \\
\end{aligned}$}
&
{$\begin{aligned}[t]
-5&.4    \\
0&.03    \\
2&.5     \\
4&.6     \\
1&.8     \\
1&.3     \\
\end{aligned}$}
\\
\midrule
Total  & \hspace*{-2.5mm} $-16$ & $10$ & $25$ & $-0.5$ \\
\bottomrule
\end{tabular}
\end{center}
\label{tab:acceptance_corrections_forward}
\end{table*}

\begin{table*}[htbp]
\caption{
Percentage variations on the differential distribution for four extreme cases of
$J/\psi$ spin alignment of maximal polarisation relative to the
nominal unpolarised assumption for $|y_{J/\psi}|<2.1$~\cite{BPHY-2014-01}.}
\begin{center}
\begin{tabular}{ c c c c c}
\toprule
$p_T^{J/\psi}$ range [GeV] & Longitudinal &  Transverse~$0$ & Transverse~$+$ &Transverse~$-$ \\
\midrule
{$\begin{aligned}[t]
(8.5&,10) \\
(10&,14) \\
(14&,18)\\
(18&,30) \\
(30&,60) \\
(60&,150) \\
\end{aligned}$}
&
{$\begin{aligned}[t]
-0&.8    \\
-4&.4    \\
-1&.9    \\
-1&.2    \\
-3&.7    \\
-1&.3    \\
\end{aligned}$}
&
{$\begin{aligned}[t]
2&.6    \\
4&.2    \\
2&.6    \\
1&.7    \\
2&.4    \\
0&.9    \\
\end{aligned}$}
&
{$\begin{aligned}[t]
39&     \\
31&     \\
21&     \\
15&     \\
8&.7    \\
4&.0    \\
\end{aligned}$}
&
{$\begin{aligned}[t]
-19&     \\
-13&     \\
-10&     \\
-8&.0    \\
-3&.1    \\
-2&.0    \\
\end{aligned}$}
\\
\midrule
Total  & $-2.3$ & $2.9$ & $28$ & $-13$ \\
\bottomrule
\end{tabular}
\end{center}
\label{tab:acceptance_corrections}
\end{table*}
 
\section{Results}

After applying the selections described above to the data, the signal is extracted and the cross-section ratio
measurement is performed in the range of $J/\psi$ transverse momentum $8.5$--$150$~\GeV\ and in two $J/\psi$
rapidity intervals, $|y_{J/\psi}|<1$ (central) and $1<|y_{J/\psi}|<2.1$ (forward).
Results are extracted in the two rapidity regions (due to the different dimuon mass resolution)
and also combined into a single rapidity range.
 
The final prompt $J/\psi + W^{\pm}$ signal yields after the application of the $J/\psi$ acceptance and muon
efficiency weights are $222\pm37$(stat) for the central region and $195\pm33$(stat) for the forward region,
where the estimated pile-up contributions are removed.
 
The total cross-section ratio is calculated for three different measurement types: fiducial, inclusive and DPS-subtracted.
The explanation of each of these methods follows, and the corresponding cross-section results are presented below
and in Tables~\ref{tab:differential_cross_section_results_dy1} and~\ref{tab:differential_cross_section_results_dy2}.
 
\subsection{Fiducial, inclusive and DPS-subtracted cross-section ratio measurements}

Due to the restrictive $\eta$ and $p_\mathrm{T}$ selection applied to the muons from the $J/\psi$,  a fiducial measurement
is made that is independent of the unknown $J/\psi$ spin-alignment or the effects of the $J/\psi$ acceptance
corrections (see Table \ref{tab:jpsiTab}) and is given by

\begin{equation*}
R^\mathrm{fid}_{J/\psi}=\frac{\sigma_\mathrm{fid}(pp\to J/\psi+W^\pm)}{\sigma(pp\to W^\pm)}\cdot\mathcal{B}(J/\psi\to\mu\mu)=\frac{1}{N(W^\pm)}
\sum_{p_\mathrm{T} \; \mathrm{bins}} [{N^{\mathrm{eff}}(J/\psi+W^\pm)}-N^{\mathrm{fid}}_{\textrm{pile-up}}]\text{ , }
\end{equation*}
 
where $N^{\mathrm{eff}}(J/\psi+W^\pm)$ is the background-subtracted yield of $W^\pm$ + prompt $J/\psi$ events after corrections
for the $J/\psi$ muon reconstruction efficiencies, $N(W^\pm)$ is the background-subtracted yield of inclusive $W^\pm$ events
and $N^{\mathrm{fid}}_{\textrm{pile-up}}$ is the expected number of pile-up background events in the fiducial $J/\psi$ acceptance.
It has been verified that the efficiency to reconstruct a $W^\pm$ is the same for the inclusive $W^{\pm}$ sample and for the
associated $J/\psi+W^{\pm}$ sample.  The result is
 
\begin{equation*}
R^\mathrm{fid}_{J/\psi}=(2.2 \pm 0.3 \pm 0.7)\times 10^{-6}\text{ , }
\end{equation*}
 
where the first uncertainty is statistical and the second is systematic.
 
The fully corrected inclusive production cross-section ratio, in which the $J/\psi$ acceptance and the unknown $J/\psi$ spin-alignment are taken into account, is given by
 
\begin{equation*}
R^\mathrm{incl}_{J/\psi}=\frac{\sigma_\mathrm{incl}(pp\to J/\psi+W^\pm)}{\sigma(pp\to W^\pm)}\cdot\mathcal{B}(J/\psi\to\mu\mu)=
\frac{1}{N(W^\pm)}\sum_{p_\mathrm{T} \; \mathrm{bins}}[N^{\mathrm{eff+acc}}(J/\psi+W^\pm)-N_{\textrm{pile-up}}]\text{ , }
\end{equation*}
 
where $N^{\mathrm{eff+acc}}(J/\psi+W^\pm)$ is the background subtracted yield of prompt $J/\psi$ + $W^\pm$  events
after $J/\psi$ acceptance corrections and efficiency corrections for the $J/\psi$ decay muons,
and $N_{\textrm{pile-up}}$ is the expected number of pile-up events in the full range of $J/\psi$ decay phase space. The result is
 
\begin{equation*}
R^\mathrm{incl}_{J/\psi}=(5.3 \pm 0.7 \pm 0.8 ^{+1.5}_{-0.7})\times 10^{-6}\text{ , }
\end{equation*}
 
where the first uncertainty is statistical, the second systematic and the third is from the spin-alignment scenario.
 
Additional measurements are made by subtracting the estimated DPS contribution in each rapidity and $p_\mathrm{T}$ interval from the inclusive cross-section ratio,
 
\begin{equation*}
R^\mathrm{DPS sub}_{J/\psi}=(3.6 \pm 0.7^{+1.1 \; +1.5}_{-1.0 \; -0.7})\times 10^{-6}\text{ , } \; \; [\sigma_{\mathrm{eff}} =15 ^{+5.8}_{-4.2} ~\mathrm{mb}]
\end{equation*}
 
and
\begin{equation*}
R^\mathrm{DPS sub}_{J/\psi}=(1.3 \pm 0.7 \pm 1.5 ^{+1.5}_{-0.7} )\times 10^{-6}\text{ , }\;\; [\sigma_{\mathrm{eff}} =6.3  \pm 1.9~\mathrm{mb}]
\end{equation*}

where the first uncertainty is statistical, the second systematic and the third is from the spin-alignment scenario.
A comparison is made with $J/\psi+W^\pm$ theory predictions, extended from the original predictions at a centre-of-mass
energy of 7~\TeV~\cite{Mao:2013xpa} to the fiducial region of this analysis at 8~\TeV\ by the same authors.
The predictions use a colour-octet long-distance matrix element (CO LDME) model for $J/\psi$ production, the parameters of
which are extracted by simultaneously fitting the differential cross-section and spin alignment of prompt $J/\psi$ production
at the Tevatron~\cite{Chao:2012iv}. These theoretical calculations include only SPS production.  They are normalised to
the $W^\pm$ boson production cross-section, calculated at next-to-next-to-leading order using the \textsc{FEWZ}
program~\cite{Anastasiou:2003ds} and corrected for the ATLAS $W^\pm$ selection requirements in Table~\ref{tab:WTab} ($5.511$~nb).
The predicted ratio is  $(0.428\pm 0.017)\times 10^{-6}$~\cite{Jpsi11,Jpsi12}.

\begin{table*}[htbp]
\caption{The fiducial and  inclusive (SPS+DPS) differential cross-section ratio in two regions of $y_{J/\psi}$.}
\begin{center}
\begin{tabular}{ c c c }
\hline
& Fiducial $[\times10^{-6}]$ & Inclusive $[\times10^{-6}]$ \\
{$y_{J/\psi}$}    & value $\pm\ \mathrm{(stat)} \pm\ \mathrm{(syst)}$ & value $\pm\ \mathrm{(stat)} \pm\ \mathrm{(syst)} \pm\  \mathrm{(spin)}$
\\
\hline
\\ [-2.0ex]
{$\begin{aligned}[t]
&|y_{J/\psi}|<1.0 \\
1.0<&|y_{J/\psi}|<2.1 \\
\end{aligned}$}
&
{$\begin{aligned}[t]
0.98  &\pm\ 0.22    &\pm\ 0&.35            \\
1.19  &\pm\ 0.25    &\pm\ 0&.35         \\
\end{aligned}$}
&
{$\begin{aligned}[t]
2.85  &\pm\ 0.52   & \pm\ 0&.44 &^{+0.87}_{-0.68}          \\
2.40  &\pm\ 0.47   & \pm\ 0&.40 &^{+0.59}_{-0.38}       \\
\end{aligned}$}
\\
\noalign{\vskip 0.5mm}
\hline
\end{tabular}
\end{center}
\label{tab:differential_cross_section_results_dy1}
\end{table*}

\begin{table*}[htbp]
\caption{The DPS-subtracted differential cross-section ratio in two regions of $y_{J/\psi}$ for two different values of $\sigma_{\mathrm{eff}}$.}
\begin{center}
\begin{tabular}{ c c c }
\hline
& DPS-subtracted  $[\times10^{-6}]$  & DPS-subtracted  $[\times10^{-6}]$ \\
& with $\sigma_{\mathrm{eff}} = 15^{+5.8}_{-4.2}$~mb  & with $\sigma_{\mathrm{eff}} = 6.3\pm1.9$~mb \\
{$y_{J/\psi}$}    & value $\pm\ \mathrm{(stat)} \pm\ \mathrm{(syst)} \pm\  \mathrm{(spin)}$ &  value $\pm\ \mathrm{(stat)} \pm\ \mathrm{(syst)} \pm\  \mathrm{(spin)}$
\\
\hline
\\ [-2.0ex]
{$\begin{aligned}[t]
&|y_{J/\psi}|<1.0 \\
1.0<&|y_{J/\psi}|<2.1 \\
\end{aligned}$}
&
{$\begin{aligned}[t]
2.05  &\pm\ 0.52    & ^{+0.54}_{-0.49} \;\;\;\;\;\;  &^{+0.87}_{-0.68}          \\
1.55  &\pm\ 0.47    & ^{+0.51}_{-0.46} \;\;\;\;\;\;  &^{+0.59}_{-0.38}       \\
\end{aligned}$}
&
{$\begin{aligned}[t]
0.94  &\pm\ 0.52   & \pm\ 0&.72 &^{+0.87}_{-0.68}          \\
0.38  &\pm\ 0.47   & \pm\ 0&.73 &^{+0.59}_{-0.38}       \\
\end{aligned}$}
 
\\
\noalign{\vskip 0.5mm}
\hline
\end{tabular}
\end{center}
\label{tab:differential_cross_section_results_dy2}
\end{table*}

\subsection{Differential production cross-section measurements}
 
\begin{figure}[h!]
\centering
\subfloat[]{
\includegraphics[width=0.45\textwidth]{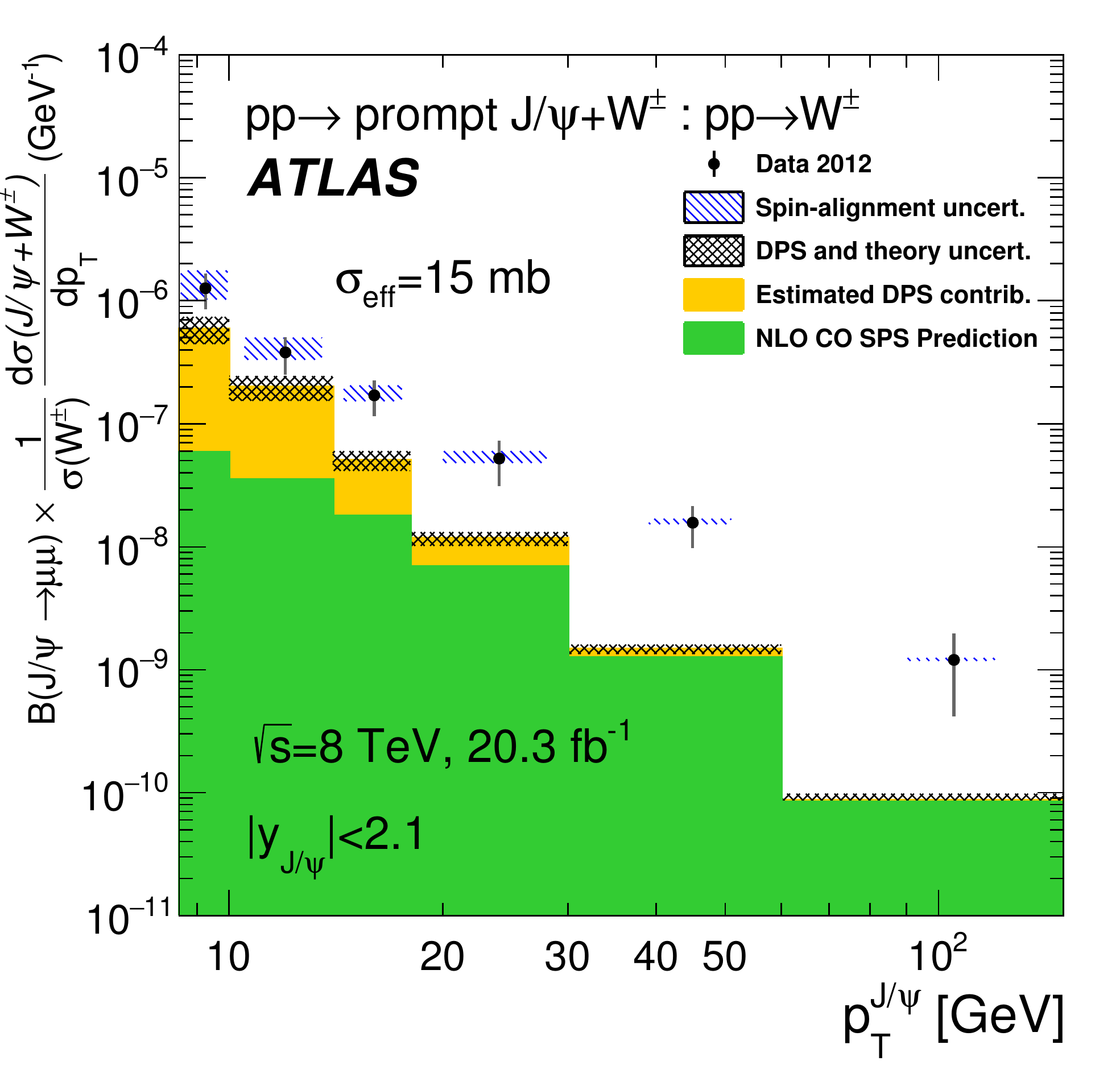}
}
\subfloat[]{
\includegraphics[width=0.45\textwidth]{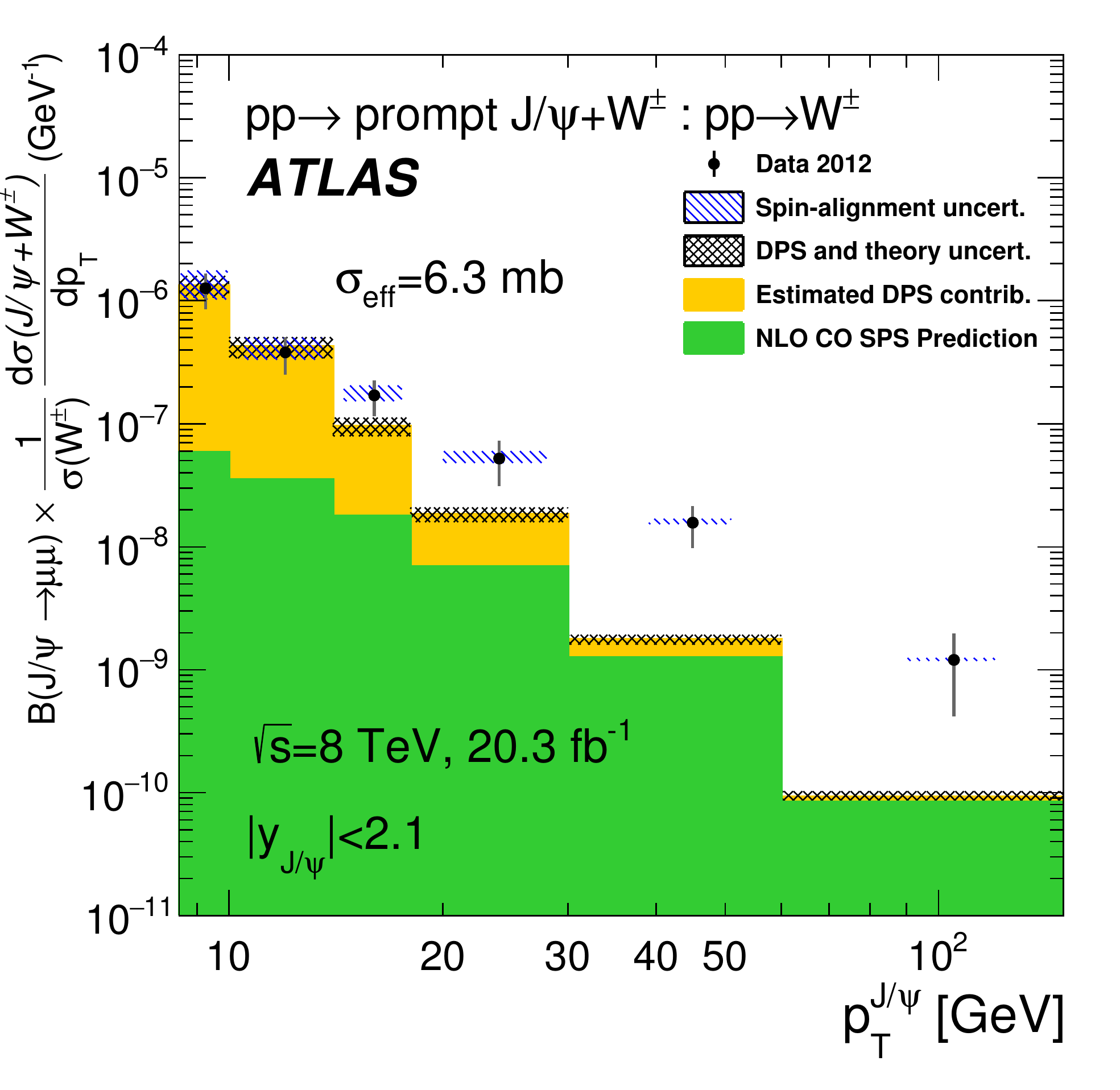}
}
\caption{The inclusive (SPS+DPS) differential cross-section ratio measurements and theory predictions
presented in six $p_\mathrm{T}^{J/\psi}$ regions for $|y_{J/\psi}| < 2.1$.
NLO colour-octet SPS predictions are shown, with LDMEs extracted from the differential cross-section and
spin alignment of prompt $J/\psi$ mesons at the
Tevatron~\cite{Mao:2013xpa,Chao:2012iv}. The DPS contribution is estimated using
(a) $\sigma_{\mathrm{eff}} = 15 ^{+5.8}_{-4.2}$~mb and (b) $\sigma_{\mathrm{eff}} = 6.3 \pm 1.9$~mb
and the method discussed in the text. The data points are identical in the two plots.}
\label{fig:crossSectionPrompt}
\end{figure}

The inclusive differential cross-section ratio, $\mathrm{d}R_{J/\psi+W^\pm}^\mathrm{incl}/\mathrm{d}p_\mathrm{T}$, is
measured for $|y_{J/\psi}| < 2.1$ in six $J/\psi$ transverse momentum intervals across the entire range of $8.5<p^{J/\psi}_\mathrm{T}<150$~\GeV,
as shown in Table \ref{tab:p_crossSection} and Figure \ref{fig:crossSectionPrompt}.
These measurements are compared with the SPS theoretical values provided by the CO model in conjuction with the estimated
DPS contribution. For $\sigma_{\mathrm{eff}} = 15$~mb, this combined prediction consistently underestimates the measurement
in all $p_\mathrm{T}$ intervals, while for $\sigma_{\mathrm{eff}} = 6.3$~mb, the summed SPS and DPS contribution
underestimates the measurement in the higher $p_\mathrm{T}$ intervals, possibly because colour-singlet processes are not included in the prediction.
 
\begin{table*}[htbp]
\caption{The measured inclusive (SPS+DPS) cross-section ratio $\mathrm{d}R_{J/\psi+W^\pm}^\mathrm{incl}/\mathrm{d}p_\mathrm{T}$ for
prompt $J/\psi$ for $|y_{J/\psi}|<2.1$. The estimated DPS contributions in each interval are listed for two possible values of $\sigma_\mathrm{eff}$.}
\begin{center}
\begin{tabular}{ l c c c}
\hline  \\[-2mm]   
$p_\mathrm{T}^{J/\psi}\,\mathrm{[\GeV]}$ & Inclusive prompt ratio $[\times10^{-7}/\,\mathrm{\GeV}]$ & \multicolumn{2}{c}{Estimated DPS $[\times10^{-7}/\,\mathrm{\GeV}]$}\\
& value $\pm\ \mathrm{(stat)} \pm\ \mathrm{(syst)} \pm\  \mathrm{(spin)}$ & $\sigma_\mathrm{eff} = 15 ^{+5.8}_{-4.2}\,\mathrm{mb}$ & $\sigma_\mathrm{eff} = 6.3\pm 1.9\,\mathrm{mb}$\\[2mm] 
\hline
\\ [-2.0ex]
{$\begin{aligned}[t]
(8.5&,10) \\
(10&,14) \\
(14&,18)\\
(18&,30) \\
(30&,60) \\
(60&,150) \\
\end{aligned}$}
&
{$\begin{aligned}[t]
12.6			&\pm\  3.3 		&\pm\ 2&.4  		& ^{+5.0}_{-2.4}       \\
3.8  		&\pm\  1.0 		&\pm\ 0&.8  		& ^{+1.2}_{-0.5}      \\
1.70  		&\pm\  0.50 		&\pm\ 0&.21  		& ^{+0.35}_{-0.17}      \\
0.52  		&\pm\  0.17 		&\pm\ 0&.12  		& ^{+0.08}_{-0.04}     \\
0.156  		&\pm\  0.054 		&\pm\ 0&.021  		& ^{+0.013}_{-0.006} 	 \\
0.012  		&\pm\  0.006 		&\pm\ 0&.005     	& ^{+0.0005}_{-0.0002} 	 \\
\end{aligned}$}
&
{$\begin{aligned}[t]
5.3 		& ^{+1.5}_{-2.1}	\\
1.64		& ^{+0.46}_{-0.64}	 	\\
0.33  		& ^{+0.09}_{-0.13}	\\
0.048		& ^{+0.013}_{-0.019}	\\
0.0021  		& ^{+0.0006}_{-0.0008}	\\
0.000032		& ^{+0.000009}_{-0.000012}	\\
\end{aligned}$}
&
{$\begin{aligned}[t]
12.7 		& \pm\ 3.8	\\  
3.9  		& \pm\ 1.2      \\  
0.77  		& \pm\ 0.23	\\  
0.114		& \pm\ 0.034	\\  
0.0049  		& \pm\ 0.0015	\\  
0.000076		& \pm\ 0.000023	\\  
\end{aligned}$}
\vspace*{2mm}
\\\hline
\end{tabular}
\end{center}
 
\label{tab:p_crossSection}
\end{table*}

\FloatBarrier

\section{Conclusion}

The ratio of the associated prompt $J/\psi$ plus $W^{\pm}$ production cross-section to the inclusive $W^{\pm}$
boson production cross-section in the same fiducial region is measured using $20.3~\mathrm{fb}^{-1}$ of proton-proton collisions
recorded by the ATLAS detector at the LHC, at a centre-of-mass energy of 8~\TeV.
The cross-section ratios are
presented for $J/\psi$ transverse momenta in the range $8.5 < p_\mathrm{T}^{J/\psi} < 150$~\GeV\ and rapidities
satisfying $|y_{J/\psi}|<2.1$.
The results are presented initially
for muons from $J/\psi$ decay in the fiducial volume of the ATLAS detector and then corrected for the kinematic
acceptance of the muons in the fiducial region.
This correction factor depends on the spin-alignment state of
the $J/\psi$ produced in association with a $W^{\pm}$ boson, which may differ from the spin alignment observed
in inclusive $J/\psi$ production. Measurements of the azimuthal angle between the $W^\pm$ boson and $J/\psi$
meson suggest that single- and double-parton-scattering contributions are both present in data. The measured
prompt $J/\psi$ + $W^{\pm}$ production rates are compared with a theoretical prediction at NLO for colour-octet
prompt production processes.   Due to the uncertainty in the value of the effective double-parton-scattering
cross-section $\sigma_{\mathrm{eff}}$, two different values are used for comparisons of theoretical predictions with data.
A smaller value of $\sigma_{\mathrm{eff}}$ brings the predicted cross-section ratio closer to the measured value; however,
neither value of $\sigma_{\mathrm{eff}}$ is able to correctly model the $J/\psi$
$p_{\mathrm{T}}$ dependence, possibly because colour-singlet processes are not included in the prediction.

\section*{Acknowledgements}
 
 
We thank CERN for the very successful operation of the LHC, as well as the
support staff from our institutions without whom ATLAS could not be
operated efficiently.
 
We acknowledge the support of ANPCyT, Argentina; YerPhI, Armenia; ARC, Australia; BMWFW and FWF, Austria; ANAS, Azerbaijan; SSTC, Belarus; CNPq and FAPESP, Brazil; NSERC, NRC and CFI, Canada; CERN; CONICYT, Chile; CAS, MOST and NSFC, China; COLCIENCIAS, Colombia; MSMT CR, MPO CR and VSC CR, Czech Republic; DNRF and DNSRC, Denmark; IN2P3-CNRS, CEA-DRF/IRFU, France; SRNSFG, Georgia; BMBF, HGF, and MPG, Germany; GSRT, Greece; RGC, Hong Kong SAR, China; ISF and Benoziyo Center, Israel; INFN, Italy; MEXT and JSPS, Japan; CNRST, Morocco; NWO, Netherlands; RCN, Norway; MNiSW and NCN, Poland; FCT, Portugal; MNE/IFA, Romania; MES of Russia and NRC KI, Russian Federation; JINR; MESTD, Serbia; MSSR, Slovakia; ARRS and MIZ\v{S}, Slovenia; DST/NRF, South Africa; MINECO, Spain; SRC and Wallenberg Foundation, Sweden; SERI, SNSF and Cantons of Bern and Geneva, Switzerland; MOST, Taiwan; TAEK, Turkey; STFC, United Kingdom; DOE and NSF, United States of America. In addition, individual groups and members have received support from BCKDF, CANARIE, CRC and Compute Canada, Canada; COST, ERC, ERDF, Horizon 2020, and Marie Sk{\l}odowska-Curie Actions, European Union; Investissements d' Avenir Labex and Idex, ANR, France; DFG and AvH Foundation, Germany; Herakleitos, Thales and Aristeia programmes co-financed by EU-ESF and the Greek NSRF, Greece; BSF-NSF and GIF, Israel; CERCA Programme Generalitat de Catalunya, Spain; The Royal Society and Leverhulme Trust, United Kingdom.
 
The crucial computing support from all WLCG partners is acknowledged gratefully, in particular from CERN, the ATLAS Tier-1 facilities at TRIUMF (Canada), NDGF (Denmark, Norway, Sweden), CC-IN2P3 (France), KIT/GridKA (Germany), INFN-CNAF (Italy), NL-T1 (Netherlands), PIC (Spain), ASGC (Taiwan), RAL (UK) and BNL (USA), the Tier-2 facilities worldwide and large non-WLCG resource providers. Major contributors of computing resources are listed in Ref.~\cite{ATL-GEN-PUB-2016-002}.
 

\printbibliography

\clearpage 
 
\begin{flushleft}
{\Large The ATLAS Collaboration}

\bigskip

M.~Aaboud$^\textrm{\scriptsize 35d}$,    
G.~Aad$^\textrm{\scriptsize 100}$,    
B.~Abbott$^\textrm{\scriptsize 127}$,    
D.C.~Abbott$^\textrm{\scriptsize 101}$,    
O.~Abdinov$^\textrm{\scriptsize 13,*}$,    
B.~Abeloos$^\textrm{\scriptsize 131}$,    
D.K.~Abhayasinghe$^\textrm{\scriptsize 92}$,    
S.H.~Abidi$^\textrm{\scriptsize 166}$,    
O.S.~AbouZeid$^\textrm{\scriptsize 40}$,    
N.L.~Abraham$^\textrm{\scriptsize 155}$
% ...
\bigskip
\\

% ...
$^{13}$Institute of Physics, Azerbaijan Academy of Sciences, Baku; Azerbaijan.\\
% ...
$^{35}$$^{(a)}$Facult\'e des Sciences Ain Chock, R\'eseau Universitaire de Physique des Hautes Energies - Universit\'e Hassan II, Casablanca;$^{(b)}$Facult\'{e} des Sciences, Universit\'{e} Ibn-Tofail, K\'{e}nitra;$^{(c)}$Facult\'e des Sciences Semlalia, Universit\'e Cadi Ayyad, LPHEA-Marrakech;$^{(d)}$Facult\'e des Sciences, Universit\'e Mohamed Premier and LPTPM, Oujda;$^{(e)}$Facult\'e des sciences, Universit\'e Mohammed V, Rabat; Morocco.\\
% ...
$^{40}$Niels Bohr Institute, University of Copenhagen, Copenhagen; Denmark.\\
% ...
$^{92}$Department of Physics, Royal Holloway University of London, Egham; United Kingdom.\\
% ...
$^{100}$CPPM, Aix-Marseille Universit\'e, CNRS/IN2P3, Marseille; France.\\
$^{101}$Department of Physics, University of Massachusetts, Amherst MA; United States of America.\\
% ...
$^{127}$Homer L. Dodge Department of Physics and Astronomy, University of Oklahoma, Norman OK; United States of America.\\
% ...
$^{131}$LAL, Universit\'e Paris-Sud, CNRS/IN2P3, Universit\'e Paris-Saclay, Orsay; France.\\
% ...
$^{155}$Department of Physics and Astronomy, University of Sussex, Brighton; United Kingdom.\\
% ...
$^{166}$Department of Physics, University of Toronto, Toronto ON; Canada.\\
% ...
$^{a}$ Also at Borough of Manhattan Community College, City University of New York, New York NY; United States of America.\\
$^{b}$ Also at Centre for High Performance Computing, CSIR Campus, Rosebank, Cape Town; South Africa.\\
$^{c}$ Also at CERN, Geneva; Switzerland.\\
$^{d}$ Also at CPPM, Aix-Marseille Universit\'e, CNRS/IN2P3, Marseille; France.\\
$^{e}$ Also at D\'epartement de Physique Nucl\'eaire et Corpusculaire, Universit\'e de Gen\`eve, Gen\`eve; Switzerland.\\
% ...
$^{*}$ Deceased

\end{flushleft}


\end{document}